\documentclass[11pt]{iopart}
\usepackage{hyperref}
\usepackage{amssymb}
\usepackage{amsopn}

\usepackage{subfigure}
\usepackage{graphicx}
\usepackage{bm}

\newcommand{\be}{\begin{equation}}
  \newcommand{\ee}{\end{equation}}
\newcommand{\ba}{\begin{eqnarray}}
  \newcommand{\ea}{\end{eqnarray}}

\renewcommand{\(}{\left(}
\renewcommand{\)}{\right)}
\renewcommand{\[}{\left[} 
\renewcommand{\]}{\right]}

  \newcommand{\lm}{\lambda}
  
  \newcommand{\PX}{P_{,X}}
  \newcommand{\half}{\frac{1}{2}}
  \newcommand{\ce}{c_e}
  \newcommand{\ca}{c_a}
  \newcommand{\de}{\delta}
  \newcommand{\De}{\Delta}
  \newcommand{\al}{\alpha}
  \newcommand{\Om}{\Omega}
  \newcommand{\dphi}{\dot\phi}
  \newcommand{\dtha}{\dot\theta}
  \newcommand{\sig}{\sigma}
  \newcommand{\qsig}{Q_{\sigma}}
  \newcommand{\qs}{Q_{S}}
  \newcommand{\s}{{\rm s}}
  \newcommand{\ds}{\delta S}
  \newcommand{\dsig}{\delta\sigma}
	\newcommand{\rd}{{\rm d}}
	\newcommand{\Ps}{P_{,S}}
	\newcommand{\PXs}{P_{,XS}}
	\newcommand{\Pss}{P_{,SS}}

\begin{document}

\title{Dynamics of entropy perturbations in assisted dark energy with mixed kinetic terms}

\author{Khamphee Karwan}
%\affiliation{
\address{
Department of Physics,\\ Faculty of Science, Kasetsart University, Bangkok 10900, Thailand \\
Thailand Center of Excellence in Physics,\\ CHE, Ministry of Education, Bangkok 10400, Thailand
}

\begin{abstract}

We study dynamics of entropy perturbations in the two-field assisted dark energy model.
Based on the scenario of assisted dark energy, in which one scalar field is subdominant compared with the other in the early epoch,
we show that the entropy perturbations in this two-field system tend to be constant on large scales in the early epoch and hence
survive until the present era for a generic evolution of both fields during the radiation and matter eras.
This behaviour of the entropy perturbations is preserved even when the fields are coupled via kinetic interaction.
Since, for assisted dark energy, the subdominant field in the early epoch becomes dominant at late time,
the entropy perturbations can significantly influence the dynamics of density perturbations in the universe.
Assuming correlations between the entropy and curvature perturbations,
the entropy perturbations can enhance the integrated Sachs-Wolfe (ISW) effect
if the signs of the contributions from entropy perturbations and curvature perturbations are opposite after the matter era,
otherwise the ISW contribution is suppressed.
For canonical scalar field the effect of entropy perturbations on ISW effect is small 
because the initial value of the entropy perturbations estimated during inflation cannot be sufficiently large.
However, in the case of k-essence, the initial value of the entropy perturbations can be large enough to affect the ISW effect to leave a significant imprint on the CMB power spectrum.

%For our choice of the metric in field space, the kinetic coupling between the fields
%can influence the effects of entropy perturbations in the case of k-essence,
%but has no  significant effect in the case of quintessence.

%For our choice of the metric in field space, the kinetic coupling between the fields
%can either reduced or enhance the effect of entropy perturbations,
%depending on the coupling strength and models of dark energy.

\vspace{3mm}
\begin{flushleft}
\end{flushleft}

\end{abstract}
%\pacs{98.80.-k}

\maketitle

\section{Introduction}

Observations, such as SNe Ia \cite{sn}, WMAP \cite{cmb}, SDSS
\cite{sdss}, etc., indicate that
the expansion of the universe is accelerating at late time \cite{Frieman:08, Chen:09}.
Within the framework of general relativity, this acceleration of the universe may be explained by
supposing that the energy density of the universe is dominated by an unknown form of energy,
whose pressure is sufficiently negative, known as dark  energy.
It has been suggested that dark energy may be in the form of cosmological constant, exotic fluid, evolving scalar fields, etc.
In general, scalar field models of dark energy, such as  quintessence\cite{Wetterich:88}- \cite{Caldwell:98} or k-essence \cite{Chiba:00,Picon:00},
may be motivated by high energy physics theory.
For example, quintessence may be a scalar field with the exponential potential
$V(\phi)=V_0 e^{-\lm \phi}$, that arises
in superstring and Kaluza-Klein theories \cite{Kanti:99}.
Here, $\lm$ is a constant parameter and the reduced Planck mass is set to be unity.

For a single quintessence field with an exponential potential \cite{Ferreira:97,Copeland:98},
  the evolution of the field in the early epoch will converge to the scaling solution for a wide range of initial conditions
if $\lm > \sqrt{3(1+w_f)}$, where $w_f$ is the equation of state parameter of the dominant fluid.
When the scaling solution is reached, the energy density of the field will evolve in the same way as that of the dominant fluid \cite{Ferreira:97,Copeland:98}.
Hence, although the scaling behaviour may provide a solution to the cosmic coincidence problem,
it cannot lead to an accelerated expansion of the universe.
The accelerated expansion of the universe can be achieved if the field evolves towards the field dominated solution with $\lm < \sqrt 2$.
Unfortunately, the range $\lm > \sqrt{3(1+w_f)}$ does not overlap with $\lm < \sqrt 2$, so that
the accelerating solution cannot be reached from the scaling solution.

In general, there is no reason why the present accelerated expansion of the universe should be driven by a single scalar field.
It has been shown that, for multiple scalar fields with exponential potentials, although the potential of each field
is too steep to drive an accelerated expansion, the cumulative dynamics of these fields can lead to an accelerated expansion of the universe \cite{Liddle:98}.
Based on this idea, multiple scalar fields with exponential potentials, with the slope of each potential  in agreement with the prediction from particle 
physics, can drive an accelerated expansion of the universe.
This idea has been originally proposed in the assisted inflationary scenario, and has also been used in dark energy model.
This model of dark energy is assisted dark energy \cite{Kim:05,Ohashi:09},
and in this model, the scaling regime can be followed by accelerated epoch.

Based on the motivation from high energy physics, it is possible that there exists an internal structure  in the field space of multiple scalar fields, e.g., in DBI models \cite{Easson:09}.
 Some features of this internal structure, which may be represented by Kinetic interactions between  fields, have been studied using a 
simple phenomenological form of kinetic interaction in several dark energy models \cite{Chimento:08} - \cite{Saridakis:10}.
It has been shown that allowing kinetic interaction in assisted dark energy can 
have only minor effects on its background dynamics \cite{Bruck:09}.

In addition to the background dynamics, the features of dark energy also depend on the dynamics of its perturbations.
In some cases, the perturbations in dark energy can leave interesting imprints on CMB anisotropies.
For example, in the case of single scalar field dark energy, the entropy (isocurvature) perturbations in dark energy can modify the ISW effect
such that the CMB power spectrum is suppressed at low multipoles,
if the  entropy perturbations in dark energy and  the curvature perturbations are initially anti-correlated
and dark energy field is nearly frozen through out the evolution of the universe \cite{GorHu:04} - \cite{kk:07}.
In this case, the dark energy field is required to be nearly frozen otherwise the entropy perturbation in dark energy damps away
before it can affect the ISW effect.
However, we will show that, for two-field  dark energy models, it is possible for the entropy perturbations 
to survive until the present epoch for a generic evolution of both fields in some case.

The dynamics of perturbations in multiple scalar field system have been studied by many authors \cite{Malik:05} - \cite{Langlois:086}.
In many cases, the dynamics of the field perturbations are conveniently described by decomposing
the field perturbations into adiabatic and entropy modes \cite{Gordon:00} - \cite{Langlois:086}.

In this work, we are interested in the behaviour of the perturbations, in particular the entropy perturbations, in the two-field assisted dark energy model.
We use the formulation in \cite{Gordon:00, Langlois:08} to decompose the perturbations in dark energy into adiabatic and entropy modes.
    We also allow  kinetic interaction between the fields in our consideration.

This paper is organized as follows.
In section 2, we present the phenomenological Lagrangian of assisted dark energy including the kinetic interaction between fields.
In section 3, we first present the evolution equations for perturbations in general two-field dark energy model,
and then apply them to the assisted dark energy model with mixed kinetic terms.
We end with the conclusion in section 4.

\section{Assisted dark energy with mixed kinetic terms}

The general action for multiple scalar field dark energy with noncanonical kinetic terms can be written as \cite{Langlois:08}
\be
S =  \int {\rm d}^4 x \sqrt{-g}\(\frac{1}{2} R  +  P(X,\varphi^I)\)\,,
\label{act1}
\ee
where index $I = 1,2, ... N$ denotes the I-th scalar field, the reduced Planck mass is set to unity and
\be
X=-\half G_{IJ} \nabla_\mu \varphi^I \nabla^\mu \varphi^J\,.
\label{def-x}
\ee
where  $G_{IJ}(\varphi^K)$ is the metric in field space and is used to raise or lower the field index $I$.
The summation convention on both field and space-time indices is implicit.
The inverse metric $G^{IJ}$ can be computed from the relation $G_{IK}G^{KJ} = \de_I^J$.
In the following, we will simplify our consideration by supposing that $G_{IJ}$ is constant.
From (\ref{act1}), one can derive the energy-momentum tensor for dark energy \cite{Langlois:08},
\be
T^{\mu \nu}=P\,  g^{\mu \nu}+\PX \, G_{IJ}\, \nabla^{\mu} \varphi^I \nabla^{\nu} \varphi^J\,,
\label{tmn}
\ee
where subscript $,_X$ denotes  derivative with respect to $X$, and the evolution equation for the scalar fields,
\be
G_{IJ} \nabla_{\mu}(\PX \nabla^{\mu} \varphi^J) + P_{,I}=0\,,
\label{kg1}
\ee
where the subscript $,_I$ denotes  derivative with respect to $\phi^I$.
In the following, we will use subscript $,_A$ to denote  derivative with respect to any variable $A$,
and use subscript $,_I$ where $I$ is the field index  to denote  derivative with respect to field $\varphi^I$.

In a spatially flat FLRW universe, the evolution equation for scalar fields takes the form
\be
\ddot \phi^I + \(3H+\frac{\dot \PX}{\PX}\)\dot \phi^I - \frac{1}{\PX}G^{IJ} P_{,J}=0\,,
\label{kg2}
\ee
where a dot denotes  derivative with respect to time $t$, $\phi^I$ is the homogeneous part of $\varphi^I$ and the Hubble parameter $H$ is given by
\be
H^2 = \(\frac{\dot a}{a}\)^2 = \frac 13 \sum_s \rho_s\,,
\label{hub}
\ee
where the index $s = r, m, d$ represents radiation, matter and dark energy respectively.
From (\ref{tmn}), the energy density of dark energy is given by
\be
\rho_d = 2 X \PX - P\,.
\label{rhod}
\ee
Multiplying (\ref{kg2}) by $\dot\phi_I$, we obtain
\be
\dot X + 6\ce^2 H X + \frac{\rho_{d,I} \dot\phi^I}{\rho_{d,X}} = 0\,,
\label{dotx}
\ee
where $\ce^2 = \PX / \rho_{d,X}$ is the effective sound speed of dark energy.

The important ingredient in  assisted dark energy model is that the
dark energy scales as the dominant fluid, i.e., undergoes the scaling regime, in the early epoch.
This implies that the equation of state parameter $w_d = P / \rho_d$ of dark energy is constant, and therefore $P \propto \rho_d$
in the scaling regime. From (\ref{hub}), one sees that $\rho_d \propto P \propto H^2 \propto t^{-2}$.
Hence, in the simplest case where $\PX$ is constant during the scaling regime, it follows from (\ref{rhod}) that $X \propto t^{-2}$,
and consequently  $\dot\phi^I \propto 1/t$ and $\phi^I \propto \ln t$,
where $\phi^I$ are the I-th fields that contribute to the dynamics of dark energy during the scaling regime.
If we further suppose that the sound speed $\ce$ is also constant during the scaling regime,
(\ref{dotx}) shows that $\rho_{d, I} \dot\phi^I \propto t^{-3}$,  which is possible when 
the $\phi^I$-dependent parts of $P$ depend on
\be
Q = \(\sum_J c_J{\rm e}^{\pm \lm_J \phi^J}\)\,,
\label{p_i}
\ee
where $c_J$ and $\lm_J$ are constant.
During the scaling regime, $\PX$ is supposed to be constant, so that 
we can write $P \propto X g(X, Q)$,
where $g(X,Q)$ and $X g_{,X}(X,Q)$ are constant during scaling regime.
A possible form of $g(X,Q)$ is $g(X,Q) = g(Y)$,
where $g(Y)$ is a polynomial function of $Y$, and $Y$ is the combination of $X$ and (\ref{p_i})
such that it is constant during scaling regime.
From the above consideration, one sees that dark energy can scale as the dominant fluid in the early epoch if its Lagrangian takes the form
\be
P(X, \phi^I) = X g(Y)\,,
\label{p-assi}
\ee
where, in general, $Y$ is the sum of the following expressions
\ba
Y_a &=& X\(\sum_J c_J{\rm e}^{\lm_J\phi^J}\),\qquad 
Y_b = \frac{X}{\sum_Jc_J{\rm e}^{-\lm_J \phi^J}}, 
\nonumber\\
Y_c &=& \frac 1X \(\sum_J c_J{\rm e}^{-\lm_J\phi^J}\), \qquad 
Y_d = \frac 1{X\(\sum_Jc_J{\rm e}^{\lm_J \phi^J}\)}\,,
\label{y}
\ea
where $\lm_J\phi^J >0$.

Let us now compute $\Om_d$ during the radiation and matter eras,
and $w_d$ during the dark energy era.
Substituting (\ref{p-assi}) into (\ref{dotx}), we get
\be
\hspace*{-2.5cm}
- 6 H \PX X = \dot X\rho_{d,X} + \rho_{d, I} \dot\phi^I
= \dot X\(2X^2g_{,XX} + 5X g_{,X} + g\) + X\(2X g_{, X I} + g_{,I}\)\dot\phi^I\,,
\label{dotx-sol}
\ee
where the index $I$ runs over the fields that contribute to the dynamics of dark energy.
It can be seen that, during radiation and matter domination, the above equation can be satisfied if $\phi^I = (2/ \lm_I)\ln t +$ constant
and hence $\dot\phi^I = (2 / \lm_I t)$.
Using this expression for $\dot\phi^I$, (\ref{p-assi}) and (\ref{y}), one can show that
\be
X g_{,X} = \frac t2 g_{,I}\dot\phi^I,
\qquad {\rm and}\qquad 
X g_{,XX} = \frac t2 g_{, X I}\dot\phi^I - g_{,X}\,.
\label{gx-gi}
\ee
Substituting the above equations into (\ref{dotx-sol}), we get
\be
\Om_d = \frac{\rho_d}{3H^2} = 3(1+w_f)\frac{\PX}{\lm_{\rm eff}^2}\,,
\label{omd}
\ee
where
\be
\frac 1{\lm_{\rm eff}^2} = G_{I I'} \frac 1{\lm_I \lm_{I'}}\,.
\label{leff}
\ee
In the above equation, the index $I'$ also runs over the fields that contribute to the dynamics of dark energy.
Similarly, one can show that $w_d$ during dark energy domination is given by
\be
w_d = -1 + \frac{\lm_{\rm eff}^2}{3\PX}\,.
\label{wdf}
\ee
It follows from (\ref{leff}) and (\ref{wdf}) that for suitable choices of $G_{I I'}$, 
$\lm_{\rm eff}$ decreases and hence $w_d$ gets closer to $-1$ when more fields contribute to the dynamics of dark energy.
This shows that the multi-field dark energy described by (\ref{p-assi}) has assisted behaviour.
Note that the form  of Lagrangian in (\ref{p-assi}) does not coincide with that given in \cite{Ohashi:09}.
This is because  the internal structure in field space as in (\ref{def-x}) can give rise to the coupling terms
between the I-th field and the other fields such as $\dot\phi^I\dot\phi_{I'}$, $\(\dot\phi^{I'}\)^2 \e^{\lm_I\phi^I}$ ($I'\neq I$) etc..
These terms are absent  in the Lagrangian in \cite{Ohashi:09}.
However, according to the above consideration, the scaling solution and assisted behaviour exist even when such terms appear in the Lagrangian.
In general,  the form of Lagrangian (\ref{p-assi}) cannot be reduced to that given in \cite{Ohashi:09} by choices in
the forms of $G_{IJ}$, $g(Y)$ or $Y$.
Nevertheless, in some cases, the both Lagrangians can become identical (see e.g. (\ref{quint})).

In the scenario for assisted dark energy, $\lm_{\rm eff}$ must be significantly larger than unity in the early epoch because
 during radiation era $\Om_d$ is required to be smaller than 0.045 according to BBN constraints \cite{Bean:01}.
For accelerated expansion of the universe, $\lm_{\rm eff}$ has to decreases at late time.
This is possible when some of the subdominant fields, that had no contributions to the dynamics of dark energy
in the early epoch, can start to contribute to the dynamics of dark energy at late time.

We now show that although (\ref{p-assi}) has assisted behaviour, 
not all the possible forms of $Y$ presented above can realize the scenario for assisted dark energy.
It can be shown that (\ref{kg2}) can be written as
\ba
\ddot\phi^I &=& - 3\ce^2 H \dot\phi^I
- \frac 12 \frac{\dot\phi^I}{X}\frac{\ce^2}{\PX}\dot\phi^K\rho_{d, K}
- \frac 12 \frac{\dot\phi^I}{X\PX}\dot\phi^K P_{,K}
+ \frac{P_{, K} G^{IK}}{\PX}\,,
\label{phid1}\\
\ddot\phi^J &=& - 3\ce^2 H \dot\phi^J
- \frac 12 \frac{\dot\phi^J}{X}\frac{\ce^2}{\PX}\dot\phi^K\rho_{d, K}
- \frac 12 \frac{\dot\phi^J}{X\PX}\dot\phi^K P_{,K}
+ \frac{P_{, K} G^{JK}}{\PX}\,.
\label{phid2}
\ea
Here, $\phi^I$ are the I-th dominant  fields that contribute to the dynamics of dark energy in the early epoch,
and $\phi^J$ are the J-th subdominant fields in the early epoch.
For simplicity, we set $G^{IJ} = \de^{IJ}$,
Substituting (\ref{p-assi}) into (\ref{phid1}) and (\ref{phid2}),
one can see that, for all possible form of $Y$, all terms on the RHS of (\ref{phid1}) are proportional to $t^{-2}$ if $\phi^I$ are in the scaling regime.
Nevertheless, for some forms of $g(Y)$ and $Y$, e.g., $g(Y) = -1 + Y = -1 + Y_a$,
the fourth term on the RHS of (\ref{phid2}) decreases with time much faster the the remaining terms
and the fourth term on the RHS of (\ref{phid1}).
We will consider this case below.
Since $\phi^J$ are subdominant, $|\dot\phi^J| \ll |\dot\phi^I|$.
According to the scenario for assisted dark energy, the expansion of the universe will start to accelerate when some of $\phi^J$ begin to be dominant.
Hence, in order to drive a present accelerated expansion of the universe, $|\dot\phi^J|$ should be comparable with $|\dot\phi^I|$ about the present epoch.
It follows from (\ref{phid1}) and (\ref{phid2}) that the ratio $|\dot\phi^J| / |\dot\phi^I|$ can increase
if $\ddot\phi^J/\dot\phi^J$ are less negative than $\ddot\phi^I/\dot\phi^I$.
For convenience, let the forth term on the RHS of (\ref{phid1}) and (\ref{phid2}) be denoted by $F_I$ and $F_J$ respectively.
In the case where $\dot\phi^J > 0$ initially, $F_I$ and $F_J$ are possitive while the remaining terms in their equations are negative.
This implies that $\ddot\phi^J/\dot\phi^J$ will be less negative than $\ddot\phi^I/\dot\phi^I$
if the initial conditions for $\phi^J$ are chosen such that $F_J / \dot\phi^J > F_I /\dot\phi^I$.
Since $F_J / \dot\phi^J$ decrease with time much faster than $F_I / \dot\phi^I$
and $F_J$ decrease faster than the remaining terms in their equation,
the ratio $|\dot\phi^J| / |\dot\phi^I|$ will increase only for a short period of time
if the initial values of $F_J/\dot\phi^J$ are not significantly larger than $F_I/\dot\phi^I$.
If the initial values of $F_J/\dot\phi^J$ increase, the  rate of growth of $|\dot\phi^J| / |\dot\phi^I|$ will  increase.
Nevertheless, the period of increasing $|\dot\phi^J| / |\dot\phi^I|$ does not expand so much.
As a result, $\phi^J$ will start to be dominant in the early time when the initial value of $F_J /\dot\phi^J$ increase.
Using  similar consideration, one will see that $\phi^J$ also cannot be dominant at the right time for the case where $\dot\phi^J <0$ initially.

In the following, we restrict ourselves to the case of two fields.
%and denote the dominant field and subdominant field in the early epoch by $\chi$ and $\xi$ respectively.
In the early epoch, the evolution of entropy perturbations in this dark energy model can be
studied analytically using (\ref{p-assi}).
However,  numerical integration is required to study the effects of entropy perturbations on CMB power spectrum,
so that we have to specify the form of $P(X,\phi^I)$.
We consider both the cases of quintessence and k-essence whose forms of $P(X,\phi^I)$ are presented in the following sections.

\subsection{The quintessence with exponential potential}

If we set $g(Y) = 1 - Y = 1 - Y_c$, (\ref{p-assi}) becomes
\be
P = X - \e^{-\lm_1 \phi^1} - \e^{-\lm_2\phi^2}\,.
\label{quint}
\ee
Here, we have set $c_1 = c_2 =1$ for convenience.
If we choose the form of $G_{IJ}$ as
\be
G_{IJ} = \de_{IJ} + \(\de_{I1}\de_{J2} + \de_{I2}\de_{J1}\) \al\,,
\label{gij}
\ee
where $|\al | < 1$ is constant,
the Lagrangian (\ref{quint}) becomes that considered in \cite{Bruck:09} and \cite{Ohashi:09} when $\al =0$.
We will use this form of $G_{IJ}$ in our numerical integration for both quintessence and k-essence models.

\subsection{The k-essence model}

We now set $g(Y) = - 1 + Y = - 1 + Y_a - M^4 Y_c$, 
where $M$ is constant, so that (\ref{p-assi}) becomes
\be
P = - X + X^2 \(\e^{\lm_1 \phi^1} + c_2\e^{\lm_2\phi^2}\) - M^4\(\e^{-\lm_1 \phi^1} + c_2 \e^{-\lm_2\phi^2}\)\,,
\label{kesse}
\ee
where we have set $c_1$ in $Y_a$ and $Y_c$ equal to unity.
The above Lagrangian has the form of the Lagrangian for dilatonic ghost condensate
model studied in \cite{Piazza:04}.
According to the previous consideration, $g(Y) = -1 + Y_a$ cannot realize the scenario for assisted dark energy
because the fourth term on the RHS of (\ref{phid2}) decreases too rapidly compared with the others.
Nevertheless, due to the term $M^4Y_c$ in the expression for $g(Y)$ in this section,
the fourth term on the RHS of (\ref{phid2}) does not decrease much faster than the remaining terms.
As a result, $\phi^2$ can contribute to the dynamics of dark energy at late time, i.e., the assisted scenario can be realized.

The value of $c_2/ M^4$ that can realize the scenario for assisted dark energy,
depends on the value of $M^4$, the starting time  for the numerical integration
and the contribution from $\phi^2$ to the dynamics of dark energy at initial stage,
i.e., the initial value of $\dot\phi^2/\dot\phi^1$ and $\e^{\lm_2\phi^2 - \lm_1\phi^1}$.
In our consideration, we set $M^4 = 1$ and $c_2 = 10^{-40}$.

\section{The perturbations in assisted dark energy}

We will mainly perform our calculations in the longitudinal gauge \cite{Bardeen:80}, in which the line element is 
\be
\rd s^2 =- (1+2\Psi)\rd t^2
+a(t)^2 (1-2\Phi)\delta_{ij}\rd x^i \rd x^j
= g_{\mu\nu}\rd x^\mu \rd x^\nu\,.
\label{long}
\ee
The evolution equation for the field perturbations $\de\phi^I = \varphi^I(t,{\rm x}) - \phi^I(t)$ can be obtained by expanding (\ref{kg1})
around the FLRW background.
From (\ref{kg1}) we write
\be
G_{IJ} g^{\mu\nu} \nabla_{\mu}(\PX \nabla_{\nu} \phi^J) + P_{,I}=0\,.
\ee
Expanding the field variable around its homogeneous background and substituting $g^{\mu\nu}$ from (\ref{long}) into the above equation,
we obtain the  following linearized equation for $\de\phi^I$ in  Fourier space:
\ba
\PX && \[\delta\ddot\phi^I + \(3H + \frac{\dot\PX}{\PX}\)\delta\dot\phi^I + \frac{k^2}{a^2}\de\phi^I\]
+ \frac 1{a^3} \(a^3\dphi^I\de \PX\)^{\displaystyle{\cdot}}
\nonumber\\
&&= \PX\dphi^I\(3\dot\Phi + \dot\Psi\) + 2 G^{IJ}P_{,J}\Psi + G^{IJ}\de P_{,J}\,,
\label{pkg}
\ea
where $\de \PX$ and $\de P_{,J}$ are the perturbations in $\PX$ and $P_{,J}$ respectively.
In the single field case this equation is in agreement with \cite{Christopherson:09, Felice:10}.
The evolution of the metric perturbations is governed by the perturbed Einstein equation \cite{Doran:03},
\ba
\fl
3H\(\dot\Phi + H \Psi\) + \frac{k^2}{a^2}\Phi &=& - \frac 12 \sum_s\de\rho_s\,,
\label{ein1}\\
\dot\Phi + H \Psi &=& \frac 12 a k^{-1} \sum_s\(1 + w_s\)\rho_s v_s,\,,
\label{ein2}\\
\Phi - \Psi &=& a^2 k^{-2}\sum_s \(\rho_s\Pi_s\)\,,
\label{ein3}
\ea
where $\de\rho_s$, $v_s$ and $\Pi_s$ are the density perturbations, velocity perturbations and anisotropic perturbations in specie $s$ respectively.
For dark energy, $\Pi_d = 0$ and
\ba
\fl
\de\rho_d = 2\de\(X\PX\) - \de P 
&=& \PX\de X + 2X\de\PX - P_{, J}\de\phi^J
= \rho_{d,X}\de X + \rho_{d, J}\de\phi^J\,,
\label{del-r}\\
(1+w_d)\rho_d v_d &=& \rho_d u_d = \PX ka^{-1} \dot\phi^J\de\phi_J\,,
\label{del-v}
\ea
where
\ba
\de X &=& \dot\phi^J\dot{\de\phi}_J - 2X\Psi\,,
\label{del-x}\\
\de P &=& \PX\de X + P_{,J}\de\phi^J\,,
\label{del-p}\\
\de\PX &=& P_{,XX}\de X + P_{,XJ}\de\phi^J\,.
\ea
Using (\ref{pkg}), (\ref{del-r}) and (\ref{del-v}), we obtain the evolution equation for $\de_d = \de\rho_d / \rho_d$ and $u_d$,
\ba
\dot\de_d &=& - 3H\(\frac{\de P}{\de\rho_d} - w_d\)\de_d + 3\(1+w_d\)\dot\Phi - \frac ka u_d\,,
\label{dot-del}\\
\dot u_d &=& H\(3w_d - 1\) u_d + \frac ka \frac{\de P}{\rho_d} + (1+w_d) \frac ka \Psi\,.
\label{dot-u}
\ea
As expected,  these equations are the evolution equations for the perturbations in a perfect fluid whose pressure perturbations are $\de P$.
These equations can be closed if the relation between $\de P$ and $\de\rho_d$ is specified.
For the the Lagrangian of the form $P(X, \varphi^I)$, it is convenient to compute this relation by
decomposing the field perturbations into adiabatic and entropy  modes.

\subsection{Decomposition into adiabatic and entropy modes}

The field perturbations can be decomposed into adiabatic and entropy modes using
a set of suitable orthonormal bases in the field space.
According to \cite{Gordon:00,Langlois:08}, we choose the first basis as
\be
\e^I = \frac{\dphi^I}{\sqrt{2 X}}\,.
\label{ei}
\ee
The other basis can be specified by solving the equations
 \be
\s^I G_{IJ} \s^J = \s_I \s^I = 1,
\qquad {\rm and} \qquad
\e^I G_{IJ} \s^J = \e^I \s_I = 0\,.
\label{si}
\ee
Using these bases, the field perturbations can be decomposed as
\be
\de\phi^I = \dsig\e^I + \ds\s^I\,.
\ee
 From the orthogonality of these bases, the adiabatic mode $\dsig$ and entropy mode $\ds$ can be written in terms of the field perturbations as
\be
\dsig = \e^I G_{IJ}\de\phi^J = \e_J\de\phi^J,,\qquad {\rm and} \qquad
\ds = \s_J \de\phi^J\,.
\ee
If this decomposition is applied to the background field, the perturbations $\de\phi^I$ should refer to a variation in time.
Hence, we can write
\be
\dot\sigma =  \e_J\dot\phi^J = \sqrt{2X},
\qquad {\rm and} \qquad
\dot S = \s_J\dot\phi^J = 0\,.
\ee
This shows that the background entropy field is constant, which is in agreement with the fact that the ``relative'' entropy
perturbations $\ds$ is gauge invariant.

Multiplying (\ref{kg2}) by $\e_I$, we obtain the evolution equation for $\sigma$
\be
\ddot\sigma = - \(3H + \frac{\dot\PX}{\PX}\)\dot\sigma + \frac{P_{,\sigma}}{\PX}
= -3\ce^2 H\dot\sigma - \frac{\rho_{d, \sigma}}{\rho_{d, X}}\,,
\ee
where $P_{,\sigma} = P_{,I}\e^I$.
On the other hand, if we multiply (\ref{kg2}) by $\s_I$, we will get a useful relation
\be
\dot\s^I = - \frac{\Ps}{\dot\sigma \PX}\e^I = - \dtha \e^I\,,
\ee
where $\Ps = P_{,I}\s^I$.
Using the relation $\e_J\s^J = 0$, we obtain
\be
\dot\e^I = \dtha\s^I\,.
\ee
The above equations can be used to compute the relation between $\de P$ and $\de\rho_d$.
The pressure perturbations $\de P$ in (\ref{del-p}) can be written in terms of $\dsig$ and $\ds$ as
\be
\de P = \PX\de X + P_{,\sigma}\dsig + \Ps\ds\,.
\ee
If we also write $\de\rho_d = \rho_{d, X}\de X + \rho_{d,\sigma}\dsig + \rho_{d, S}\ds$ and use
the relation $\PX = \ce^2\rho_{d, X}$, we will get
\be
\de P = \ce^2 \de\rho_d + \(P_{,\sigma} - \ce^2\rho_{d, \sigma}\)\dsig
+ \(P_{,S} - \ce^2\rho_{d, S}\)\ds\,.
\label{dp0}
\ee
Since $\dot S = 0$, $P_{,\sigma} - \ce^2\rho_{d, \sigma} = \(\dot P - \ce^2\dot\rho_d\)/\dot\sigma$.
Using the definition of the adiabatic sound speed $\ca^2 = \dot P / \dot\rho_d$, and 
the conservation equation for $\rho_d$, $\dot\rho_d = - 3H\(1+w_d\)\rho_d$,
the second term on the RHS of the above equation becomes
\be
3H\(\ce^2 - \ca^2\)\PX\dot\sigma\dsig = 3 H \frac ak\(\ce^2 - \ca^2\)\rho_d u_d\,.
\ee
Here, we have written $u_d$ in terms of the adiabatic field as
\be
\rho_d u_d = \PX ka^{-1} \dot\phi^J\de\phi_J =  \PX ka^{-1} \dot\sigma\dsig\,.
\ee
From (\ref{rhod}), we can write the third term on the RHS of (\ref{dp0}) in terms of the pressure only.
Finally, we have
\be
\de P = \ce^2\de\rho_d + 3 \frac ak H \(\ce^2 - \ca^2\)\rho_d u_d + \[\Ps\(1 + \ce^2\) - 2\ce^2 X \PXs\]\ds\,.
\label{dp-ds}
\ee
If $\ds = 0$, the form of this relation will be the same as the one for single scalar field derived in \cite{GorHu:04}.
This relation implies that (\ref{dot-del}) and (\ref{dot-u}) can be closed if the evolution of $\ds$ is known.

In order to derive the evolution equation for $\ds$, the following relations \cite{Langlois:08} are required,
\ba
\dot \Ps &=& -\frac{\Ps P_{,\sigma}}{\dot\sigma \PX}+P_{,\sigma S}\dot\sigma+\PXs\dot\sigma\ddot\sigma\,,
\\
P_{,\sigma S} &=&  P_{,IJ}\e^I\s^J, \qquad {\rm and} \qquad \PXs = P_{,X J}\s^j\,.
\ea
The evolution equation for $\ds$ is computed by multiplying (\ref{pkg}) with $\s_I$.
The result is
\be
\ddot{\ds} + \(3H + \frac{\dot{\PX}}{\PX}\)\dot{\ds} + \(\frac{k^2}{a^2} + \mu_S^2 + \frac{\Xi^2}{\ce^2}\)\ds
= - \frac{\Xi}{\dot\sigma \PX}\de\rho_{dv}\,,
\label{dot-s}
\ee
where
\ba
\mu_S^2 &\equiv & -\frac{\Pss}{\PX} - \frac{1}{2 \ce^2 X}\frac{\Ps^2}{\PX^2}+2\frac{\PXs \Ps}{\PX^2}\,,
\\
\Xi &\equiv & \frac{1}{\dot \sigma\PX}\[(1+\ce^2)\Ps - \ce^2 \PXs\dot\sigma^2\]\,,
\ea
and $\de\rho_{dv}$ is the density perturbations of dark energy in the dark energy rest frame, given by
\be
\hspace*{-2.5cm}
\de\rho_{dv} = \de\rho_d + 3 H \frac ak \rho_d u_d
= \rho_{d, X}\(\dot\sigma\dot\dsig - 2X\Psi - \ddot\sigma\dsig\) - \[\Ps \(\frac 1{\ce^2} + 1\) - 2X\PXs\]\ds\,.
\ee
In the second equality, the expression of $\de\rho_d$ in terms of the adiabatic and entropy fields,
\begin{eqnarray*}
\hspace*{-2.5cm}
\de\rho_d = \rho_{d,X}\de X + \rho_{d, J}\de\phi^J
= \rho_{d, X}\(\dot\sigma\dot\dsig - 2X\Psi \) + \rho_{d, \sigma}\dsig - \[\Ps \(\frac 1{\ce^2} + 1\) - 2X\PXs\]\ds\,,
\end{eqnarray*}
has been used.
For the early epoch, it is possible to solve
(\ref{dot-del}), (\ref{dot-u}) and (\ref{dot-s}), which completely describe the dynamics of perturbations in dark energy, analytically 
Nevertheless,  numerical  integration is required to study the behavior of perturbations in dark energy about the present epoch.
In order to solve these equations numerically, the initial conditions for $\de_d$, $u_d$, $\ds$ and $\dot{\ds}$
in the early epoch must be specified.
Since the contribution from entropy perturbations $\ds$ is small in the early era,
it is possible to use the following adiabatic initial conditions for $\de_d$ and $u_d$ if $w_d$ is approximately constant during the initial stage.
\be
\frac{\de_d}{1 + w_d} = \frac{\de_r}{1 + w_r},\qquad {\rm and}\qquad \frac{u_d}{1+ w_d} = v_r\,,
\label{adia-con}
\ee
where $\de_r = \de\rho_r/\rho_r$ and $v_r$ is the density contrast and velocity perturbations of radiation respectively.
Supposing that dark energy is a fundamental  field which has already existed since inflationary era,
the initial magnitude of $\ds$ may be estimated by studying the dynamics of perturbations in dark energy during inflation.
 
\subsection{The initial conditions from inflation}

Here, we would like to roughly estimate the initial magnitude of $\ds$ by studying the evolution of perturbations in dark energy during inflation.
In the following calculation, we assume for simplicity that inflation is driven by a single canonical inflaton field, the energy density of dark energy is subdominant,
dark energy and inflaton are weakly coupled via  gravity only.
Since it is convenient to study the dynamics of perturbations during inflation using spatially flat gauge,
we write the evolution equation for perturbations (\ref{pkg}) 
in spatially flat gauge using the relations \cite{Malik:01, Riotto:02}
\be
\Psi_{\rm flat} = \Psi_{\rm long} + \(\frac{\Phi_{\rm long}}{H}\)^{\displaystyle{\cdot}}\,,
\qquad
Q^I = \de\phi_{\rm flat}^I = \de\phi_{\rm long}^I + \dot\phi^I\frac{\Phi_{\rm long}}{H}\,.
\ee
where subscripts ``flat'' and ``long'' denote spatially flat and longitudinal gauge respectively.
After performing a transformation , the metric perturbations $\Psi_{\rm flat}$ can be eliminated using the relation \cite{Gordon:00}
\be
\Psi_{\rm flat} = \frac{1}{2H}\dot\psi Q_f\,,
\label{psiflat}
\ee
where $\psi$ is the background inflaton field
and $Q_f$ is the perturbed inflaton field in spatially flat gauge.
It is not hard to show that (\ref{pkg}) in spatially flat gauge
is compatible with the action
\ba
\hspace*{-2cm}
%\fl
S^{(2)} &=& \half \int {\rm d}t\, {\rm d}^3{\rm k} \,a^3
\[\(\PX G_{IJ} + P_{,XX} \dphi_I \dphi_J \)\dot Q^I \dot Q^J
-\frac{\PX k^2}{a^2}G_{IJ}Q^I Q^J 
\right.
\nonumber\\
\hspace*{-2cm}
&&  
  \left.
  + P_{,IJ}Q^I Q^J - C_K Q_f Q^K
- D_K\dot Q_f Q^K
+ 2 P_{,XJ} \dphi_I Q^J \dot Q^I\]\,,
\label{flat1}
\ea
whose form is similar to that in \cite{Langlois:08},
except for the third, fourth and fifth terms in the action.
This is because the metric perturbations $\Psi_{\rm flat}$ in (\ref{psiflat})
are sourced by the inflaton field not by $Q^I$ in our case.
Thus, to simplify our task, we follow the calculation in \cite{Langlois:08}.
Here, inflaton and dark energy are coupled via the perturbed metric, and the coupling coefficients are given by 
\be
\hspace*{-2cm}
C_K = \frac{2X}{H} P_{,XK}\dot\psi
-\frac{1}{a^3}\[\frac{a^3}{H}\PX\(1+\frac{1}{\ce^2}\)\dot\psi\dot\phi_K\]^{\displaystyle{\cdot}}\,
\qquad
D_K =\frac{\PX}{H}\(1 - \frac 1{\ce^2}\)\dot\phi_K\dot\psi\,.
\label{c-inflaton}
\ee
The action (\ref{flat1}) can be written in terms of the adiabatic and entropy fields as
\ba
\hspace*{-2cm}
 S^{(2)} 
&=&   \half \int  {\rm d}t\, {\rm d}^3{\rm k} \,a^3\
\[ \frac{\PX}{\ce^2}\(\dot\qsig - \dtha \qs\)^2 +\PX \(\dot \qs + \dtha\qsig\)^2
- \frac{\PX k^2}{a^2}Q_MQ^N
 \right.\nonumber\\
\hspace*{-2cm}
  &+&\left. 
P_{, MN} Q^MQ^N - C_M Q_f Q^M - D_M \dot{Q}_f Q^M
+2\dot\sig \(\dot\qsig - \dtha \qs\)P_{,XM}Q^M\]\,,
\label{flat2}
\ea
where indices $M, N$ run over $\sig, S$.
Using the new fields which are canonically normalized,
\be
v_{\sig}=\frac{a \sqrt{\PX}}{\ce}\, \qsig = \beta \qsig\,,\qquad \,v_S = a\,\sqrt{\PX}\, \qs = \gamma\qs\,,
%\qquad v_f = a Q_f,
\label{v}
\ee
together with conformal time $\tau = \int d a /a$, the second-order action (\ref{flat2}) can be rewritten in the form
\ba
\label{flat3}
\hspace*{-1.9cm}
S^{(2)}=\frac{1}{2}\int {\rm d}\tau {\rm d}^3{\rm k} 
&&\[ 
  v_\sig^{\prime\, 2}+ v_S^{\prime\, 2} -2\xi v_\sig^\prime v_S-k^2 \ce^2 v_\sig^2 -k^2 v_S^2 
\right. \\
\hspace*{-1.9cm}
&& \left.
+\Omega_{\sig\sig}v_\sig^2+\Omega_{SS} v_S^2+2\Omega_{\sig S}v_\sig v_S
+ \bar{C}_\sig Q_f v_\sig + \bar{C}_S Q_f v_S
+ \bar{D}_\sig Q'_f v_\sigma
\]\,,
\nonumber
\ea
where a prime denotes  derivative with respect to conformal time, and
\ba
\hspace*{-2.7cm}
\xi&=& \frac{a}{\ce}\Xi=\frac{a}{\dot\sig \PX \ce}[(1+\ce^2)\Ps - \ce^2 \dot \sig^2 \PXs],
\nonumber\\
\hspace*{-2.7cm}
\Omega_{\sig\sig} &=& \frac{\beta''}{\beta} + a^2\ce^2\dtha^2 + \frac{a^2\ce^2}{\PX}\(P_{,\sig\sig} - \(a^3 \dot\sig P_{,X\sig}\)^{\displaystyle{\cdot}}\)\,,
\quad  
\Omega_{\sig S}= \(\frac{\beta'}{\beta} + a\frac{\ddot\sig}{\dot\sig}\)\xi\,, 
\quad
\Omega_{SS}=\frac{\gamma''}{\gamma}-a^2 \mu_S^2\,,
\nonumber\\
\hspace*{-2.7cm}
\bar{C}_\sig &=&  \frac{a^3 \ce}{\sqrt{\PX}}\[
\[\(1+\frac{1}{\ce^2}\)\frac{\dot\psi}{H}\]^{\displaystyle{\cdot}} P_{,\sig}
- \frac{2X}{H} P_{,X\sig}\dot\psi\],
\quad
\bar{C}_S =
\frac{a^2\dot\sig \sqrt{\PX}}{\ce}\frac{\dot\psi}{H}\xi,
%\frac{a^3}{\sqrt{\PX}}\[\(1+\frac{1}{\ce^2}\)\frac{\dot\psi}{H} \Ps
%- \frac{2X}{H} \PXs\dot\psi\].
\nonumber\\
\hspace*{-2.7cm}
\bar{D}_\sig &=& \frac{a^2\dot\sig}{\PX}\(\frac{1}{\ce} - \ce\)\frac{\dot\psi}{H}\,.
\ea
The evolution equation for entropy mode can be derived from (\ref{flat3}) as
\be
\hspace*{-1cm}
v_S''+\xi  v_\sig'+\(k^2- \frac{\gamma''}{\gamma}+a^2\mu_S^2\) v_S - \(\frac{\beta'}{\beta} + a\frac{\ddot\sig}{\dot\sig}\) \xi v_\sig = \frac{a^2\dot\sig \sqrt{\PX}}{2\ce}\frac{\dot\psi}{H}\xi Q_f\,.
\label{dvs}
\ee
In order to solve this equation, the terms that are propportional to $\xi$ and $\mu_S^2$ are supposed to be negligible, i.e.,
$|\xi | \ll aH$ and $\mu_S^2 \ll H^2$, which is possible when one dark energy field is subdominant compared with the other.
We will discuss this issue in the next section.
According to the scenario for assisted dark energy, the field, which undergoes scaling regime during radiation era,
has larger contribution to the energy density as compared to the other before it enters the scaling regime.
Since, before the scaling regime starts, the subdominant field tends to be nearly frozen \cite{Ohashi:09}
while the dominant one can undergo a long period of kinetic regime (see figure (\ref{fig:1})),
the contribution of the dominant field to the energy density of dark energy does not tend to decrease when looking back in time.
Therefore, it is possible that the field that is dominant during radiation era is also dominant during inflation.
 If $|\xi | \ll a H$, $\mu_S^2\ll H^2$ and the dark energy fields slowly evolve, i.e., $\PX$ and $\ce$
evolve slower than the scale factor, the approximate solution of (\ref{dvs}) is
\be
v_S(k) \approx  \frac{1}{\sqrt{2k}}e^{-ik \tau }\(1-\frac{i}{ k \tau}\)\,.
\label{vsk}
\ee
For a slow rolling inflaton, it can be shown that \cite{Riotto:02}
\be
|Q_f| \approx \frac 1{a\sqrt{2 k}}\,,
\ee
on large scales, so that
\be
\hspace*{-2cm}
||\ds || = ||\qs || \simeq \frac 1{\sqrt{\PX}}||Q_f || = \frac 1{\sqrt{\PX}}\frac{\dot\psi}{H}||{\cal R}|| = \frac 1{\sqrt{ \PX}} \sqrt{2\epsilon} ||{\cal R}||
\simeq \frac 1{\sqrt{ \PX}} \sqrt{0.125 R} ||{\cal R}||\,,
\label{init-ds}
\ee
where ${\cal R}$ is the curvature perturbations on the comoving hypersurfaces, $\epsilon$ is the slow roll parameter of inflaton,
$R$ is the relative amplitude  of the tensor to  scalar perturbations,
and $||\cdots ||$ denotes $\sqrt{\langle \(\cdots\)^2\rangle}$ and $|\cdots |$
when $\qs$ and $Q_f$ are uncorrelated and fully correlated respectively.
Here, $\langle\cdots\rangle$ refers to the ensemble average.
In the second equality, we have used the fact that $\ds$ is gauge invariant.
It follows from this equation that the magnitude of $\ds$ can be large if $|\PX |$ is small,
which is possible when dark energy slowly evolves in many k-essence models,
e.g., in the case of Lagrangian in (\ref{kesse}),
$\PX \rightarrow 0$ when $w_d \rightarrow -1$ if $M^4 = 1$.
This result can be used to estimate the initial magnitude of $\ds$ during the radiation era,
because, as we will see in the next section, $\ds$ tends to be constant for generic evolution of dark energy as long as one of the scalar fields is subdominant.

Since we use the Lagrangian of the form $P(X, \varphi^I)$ for dark energy in the above calculation,
the adiabatic and entropy mode of the dark energy perturbations propagates with the speed of sound $\ce$ and speed of light respectively.
However, in more general case in which the Lagrangian takes the form $P(X^{IJ}, \varphi^K)$ \cite{Langlois:086}
where $X^{IJ}= -(1/2)\nabla_\mu \varphi^I \nabla^\mu \varphi^J$,
the propagation speed of adiabatic and entropy modes can be different from the ones considered here.
For example,  in the case of DBI Lagrangian \cite{Langlois:084, Langlois:086},
both adiabatic and entropy modes propagate with the speed of sound, so that
the gradient term in the second order perturbed action is multiplied by sound speed.
In this case, the amplitude of the entropy field $\ds$ can be amplified compared with the adiabatic field if $|\ce| < 1$.
However, in our case, it follows from (\ref{init-ds}) that the magnitude of $\ds$ can be enhanced compared with the curvature perturbation 
when $|\PX | < 1$, because, in order to obtain the canonically normalized field variable,
we rescale the entropy field $\ds$ by the scale factor and square root of is kinetic coefficient $\sqrt{\PX}$,
and the curvature perturbation is generated by canonical inflaton field.

In general, it is not convenient to use the value of $\qsig$ during inflation to define the initial conditions for $\dsig$ in the radiation era,
because the evolution of $\qsig$ depends on the evolution of dark energy after inflation.
Hence, we define the initial conditions for the $\dsig$ and also $\dot{\dsig}$ through 
the adiabatic conditions for the perturbations in the total energy density and velocity given in (\ref{adia-con}).
However, strictly speaking, (\ref{adia-con}) can be justified if the curvature perturbation and the perturbations in dark energy have correlation.
Nevertheless, in the above calculation, the perturbations in dark energy and curvature perturbation are uncorrelated
because we suppose that $|\xi| \ll a H$, $\mu_S^2\ll H^2$ and the inflaton slowly evolves.
Several models for generating correlated dark energy perturbations have been proposed \cite{GorHu:04, Moroi:04}.
To simplify our task, instead of considering these models in detail, we use the model in \cite{Moroi:04} to estimate
how (\ref{init-ds}) is modified when the perturbations in dark energy and curvature perturbation are correlated.
Based on the model in \cite{Moroi:04}, the correlated dark energy perturbations can be generated
if the curvaton field is coupled with the dark energy fields such as $g_I\varphi^I \zeta$
where $\zeta$ is the curvaton field and $g_I$ is the coupling coefficient.
In the case where $|\xi| \ll a H$, $\mu_S^2\ll H^2$ and $\PX$ together with $\ce$ evolves slower than the scale factor,
the dynamics of $\qs$ and $\qsig$ should not be much different from the dynamics of perturbations in dark energy field in  \cite{Moroi:04}.
Hence, comparing the result in \cite{Moroi:04} with (\ref{init-ds}),
one expects that (\ref{init-ds}) are modified only by multiplication factor whose magnitude is smaller than unity,
such as $|\ds| < \sqrt{0.125 R}|{\cal R}|/\sqrt{\PX}$,
 when the entropy perturbations in dark energy and curvature perturbation are correlated.
Thus, in the following consideration, we assume that the perturbations in dark energy and curvature perturbation are fully correlated
and use the RHS of the last equality in (\ref{init-ds}) to give a rough estimate of the upper bound of $|\ds|$.

%However, in \cite{}, the authors show that the ratio of the
%correlated isocurvature (entropy) perturbations in dark energy field to the curvature perturbation depends on $1/Z$
%where $Z$ is the kinetic coeficient of the perturbed dark energy field.
%This result arises because, in order to study the quantum fluctuations in scalar fields during inflation,
%we redefine the fields such that the perturbed fields in the action have a canonical kinetic term, similar to the above calculation.
%This result is roughly similar to (\ref{init-ds}).
%Hence, in the following consideration, we will assume that the perturbations in dark energy and curvature perturbation are (fully) correlated
%and (\ref{init-ds}) is approximately valid.

\subsection{Evolution of the entropy perturbations	 during the radiation and matter epochs}

In order to study the evolution of the entropy perturbations,
the form of $\s^I$ should be known.
In the case of two dimensional field space, applying the constraint equations for basis $\s^I$ in (\ref{si}) to a symmetric $G_{IJ}$ yields
\be
\s^1 = \mp \frac{G_{22}\e^2 + G_{12}\e^1}{\sqrt{\det G}}
\qquad {\rm and} \qquad 
\s^2 = \pm \frac{G_{11}\e^1 + G_{12}\e^2}{\sqrt{\det G}}\,,
\label{si-gen}
\ee
where $\det G > 0$.
Obviously, there are two possible choices of $\s^I$ in the two dimensional field space.
These choices correspond to two $\s^I$ pointing in opposite directions.
Nevertheless, these different choices of $\s^I$ do not lead to  different dynamics of perturbations in dark energy.
This is because $\ds$ and the derivative $\partial^n /\partial S^n$, where $n$ is an odd integer, change sign for these different choices of $\s^I$,
so that the evolution equation for $\ds$ in (\ref{dot-s}), the relation between $\de P$ and $\de\rho_d$
in (\ref{dp-ds}) and the expression for $\de\rho_d$ in terms of adiabatic and entropy modes are the same for these different choices.
However, it follows from (\ref{init-ds}) that
these different choices of $\s^I$ imply a change in the relative sign between $\ds$ and the curvature perturbations.
Nevertheless, this relative sign cannot be computed explicitly from the choices of $\s^I$.
In our consideration, we choose $\s^1 = - (G_{22}\e^2 + G_{12}\e^1)/\sqrt{\det G}$
and $\s^2 = (G_{11}\e^1 + G_{12}\e^2)/\sqrt{\det G}$,
and set the ratio $\ds / {\cal R}$ to be negative initially,
i.e., $\ds$ and the curvature perturbation are anti-correlated.
We set $\ds / {\cal R}< 0$ initially because it leads to
the suppression of the CMB power spectrum at low multipoles for our choice of Lagrangians (\ref{quint})
and (\ref{kesse}) in our numerical integration.
However, in general, initial $\ds / {\cal R}< 0$  does not necessarily lead to the suppression of the 
CMB power spectrum at low multipoles.

In the radiation and matter epochs, when the field $\phi^2$ is subdominant compared with $\phi^1$,
the magnitude of $\s^I$ can be estimated by solving (\ref{kg2}).
Rewriting (\ref{kg2}) as
\be
\frac{d}{d t}\(a^3\PX \dot\phi^I\) = a^3 G^{IJ}P_{, J}\,.
\ee
When $\phi^2$ is subdominant, $|P_{,1}| \gg |P_{,2}|$, and hence
\be
\hspace*{-1.5cm}
\dot\phi^1 = \frac{G^{11}}{a^3\PX}\int dt \(a^3 P_{,1}\) + \frac{c_1}{a^3\PX}\,,
\qquad
\dot\phi^2 = \frac{G^{12}}{a^3\PX}\int dt \(a^3 P_{,1}\) + \frac{c_2}{a^3\PX}\,,
\ee
where $c_1$ and $c_2$ are the constants of integration.
This result implies that
\be
\dot\phi^2 = - \frac{G_{12}}{G_{22}}\dot\phi^1 + \frac 1{a^3\PX}\(c_2 + \frac{G_{12}}{G_{22}}c_1\)\,.
\label{p2p1}
\ee
Here, we have used
\be
G^{12} =  - \frac{G_{12}}{\det G}
\quad
{\rm and}
\quad
G^{11} = \frac{G_{22}}{\det G}\,.
\ee 
Substituting (\ref{p2p1}) into (\ref{def-x}), we get
\be
X = - \frac 12 \(\dot\phi^1\)^2 \(G_{11} - \frac{G_{12}^2}{G_{22}}\) - \frac{G_{22}}2 \[\frac 1{a^3\PX}\(c_2 + \frac{G_{12}}{G_{22}}\)\]^2\,.
\label{x-ear}
\ee
In the case of our interest where $-1 < w_d <1$, the first term on the RHS of this equation will not decrease faster than the second term,
so that the second term will be negligible compared with the first term as the universe evolves.
Thus, we ignore this term in the following calculation.
%The vanishing of the second term is required for $\phi^1$ to be in the scaling regime.
%Moreover, the assumption that $\phi^2$ is subdominant and hence does not alter
%the dynamics of dark energy implies that $X \propto \(\dot\phi^1\)^2$, so that
%the integration constants $c_1$ and $c_2$ should be ignored.
%Note that in the case of dilatonic ghost condensate, the case where $1/3 < w_d < 1$ is excluded, because $\rho_d$ is not real and positive in this case.
Substituting (\ref{p2p1}) into (\ref{si-gen}), we obtain
\be
|\s^1| \ll |\s^2| \simeq \left |\frac{\sqrt{\det G}}{G_{22}}\e^1\right |\,,
\qquad {\rm where}\qquad 
|\e^1| \simeq \left |\sqrt{\frac{G_{22}}{\det G}}\right | \gg |\e^2|\,.
\label{s1ll}
\ee
Here, the expression for $|\e^1|$ is computed using (\ref{x-ear}) and (\ref{ei}).
In the following calculations, we suppose that $|\s^2| \sim |\e^1| \sim {\cal O}(1)$.

Let us now come to the case of assisted dark energy whose Lagrangian is given by (\ref{p-assi}).
Using this Lagrangian, we estimate the magnitudes of $\mu_S^2$ and $\Xi$ which
correspond to the amplitude of the ``effective mass'' of $\ds$ and the coupling between $\ds$ and $\de\rho_d$.
From (\ref{gx-gi}),
one can derive the following relations in the case where $\phi^2$ is subdominant:
\be
%\hspace*{-1.7cm}
%\lm_1 |g_{,1}| \sim |g_{,11}| \gg \lm_1|g_{,2}|, |g_{, 12}|\,\,\, {\rm and}\,\,\, |g_{,22}|\,,
|\lm_1| X g_{, X} \simeq g_{,1}\,,
\qquad {\rm and} \qquad
%\lm_1 X g_{, X 1} \simeq g_{,11}
|\lm_1 g_{,1}| \sim |g_{,11}| \,.
\label{xgx2g1}
\ee
Using these relations, $|P_{,1}| \gg |P_{,2}|$ and (\ref{s1ll}), if $\lm_1$ is not very much larger than unity, it can be seen that
\ba
\hspace*{-2.2cm}
X |\PXs| &=& X |\(g + X g_{,X}\)_{,I}\s^I| \simeq X \left |\(g + \frac{g_{,1}}{\lm_1}\)_{,I}\s^I\right | \sim |\Ps|\,.
\label{pxs-ps}\\
\hspace*{-2.2cm}
|P_{,SS}| &=& |\s^1\s^1 P_{,11} + 2\s^1\s^2 P_{,12} + \s^2\s^2 P_{,22}| = |P_{,11}B_{SS}| \sim |\lm_1 P_{,1}B_{SS}| \simeq |\lm_1 P_{,\sig}B_{SS}|,
\label{pss-psig}
\ea
where $B_{SS} \equiv \s^1\s^1 + 2\s^1\s^2P_{,12}/P_{,11} + \s^2\s^2P_{,22}/P_{,11}$
so that $|B_{SS}|\ll 1$ when $\phi^2$ is subdominant.
The relation (\ref{pxs-ps}) is valid when $\PXs \neq 0$.
From $P_{,1}\gg P_{,2} and $(\ref{s1ll}), it is easy to see that
\be
|\Ps | = |P_{,1}\s^1 + P_{,2}\s^2| = |P_{,1}B_S| \simeq |P_{,\sig}B_S|\,,
\label{ps-psig}
\ee
where $ B_S \equiv \s^1 + \s^2P_{,2}/P_{,1}$
so that $|B_S|\ll 1$ when $\phi^2$ is subdominant.
Using (\ref{xgx2g1}), (\ref{p-assi}) and $(1+w_d)\rho_d = 2X\PX$, we get
\be
\hspace*{-2cm}
\frac{P_{,\sig}}{\rho_d} = \frac{X g_{,\sig}}{2X^2 g_{,X} + X g} \simeq \frac{\lm_1 X g_{,X}}{2 X g_{,X} + g} \simeq (1 - w_d)\frac{\lm_1}2\,,
%\leq \lm_1\,,
\quad {\rm and}\quad
\frac{P_{,\sig}}{X\PX} \simeq \lm_1\frac{1 - w_d}{1 + w_d}\,.
\label{psig-rd}
\ee
Hence, $|\Ps |/ \rho_d \simeq \lm_1(1 - w_d)|B_S|/2 \ll 1$,
implying that the contribution of $\ds$ to $\de P$ is negligible in the early epoch.
In the early epoch, when the contribution from $\phi^2$ to $\rho_d$ is significantly small such that $|B_{SS}| \ll 1+w_d$
and $|B_S| \ll 1+w_d$, it follows from (\ref{pss-psig}), (\ref{ps-psig}) and (\ref{psig-rd}) that
\ba
\left |\frac{\Ps}{\dot\sig \PX \sqrt{\rho_d}}\right | &\lesssim & \left |\frac{P_{,\sig}B_S}{X \PX}\right | 
\simeq (1 - w_d)\frac{|\lm_1 B_S|}{1+w_d} \ll 1\,,
%\qquad {\rm and} \qquad
\label{ps-px}\\
\left |\frac{P_{,SS}}{\rho_d\PX}\right | &\lesssim & \left |\frac{\lm_1P_{,\sig}B_{SS}}{X\PX}\right |
 \simeq (1 - w_d)\frac{\lm_1^2 |B_{SS}|}{1+w_d} \ll 1\,.
\label{pss-px}
\ea
In the above calculations, we have used $|\dot\sig | \lesssim \sqrt{\rho_d}$ instead of $|\dot\sig| \sim \sqrt{\rho_d}$,
because $|\dot\sig|$ can be much smaller than $\sqrt{\rho_d}$ when $w_d\rightarrow -1$ in some cases, e.g.,
in the case of quintessence models where $P = X - V$.
Using (\ref{pxs-ps}), (\ref{ps-px}), (\ref{pss-px})
 and $\sqrt{\rho_d} < H$ together with $\ce^2 \sim {\cal O}(1)$,
it can be shown that $|\Xi | \ll H$ and $\mu_S^2 \ll H^2$.

When $|\Xi | \ll H$ and $\mu_S^2 \ll H^2$, the evolution equation for $\ds$ (\ref{dot-s}) on super horizon scales becomes
\be
\frac{d}{d t} \(a^3\PX\dot{\ds}\) \approx 0\,.
\ee
The solution of this equation is
\be
\dot{\ds} = \frac{d_1}{a^3\PX}\,,
\qquad
\ds = \int \frac{d_1 dt}{a^3\PX} + {\rm const.} \sim - \frac{\dot{\ds}}{H} + {\rm const}\,,
\label{ds-long}
\ee
where $d_1$ is the constant of integration.
To evaluate the integral in the above expression  for $\ds$, $\PX$ is assumed to be approximately constant.
Since, according to the above equations, $|\dot{\ds} / H|$ decreases with time,
and $|\dot{\ds} / H| < |\ds|$ on large scales during inflation because the dark energy fields slowly evolve,
we get $\ds \simeq$ constant on large scales after the inflationary era.
%Using $H a\sim 1 / \tau$ during inflation and differentiating (\ref{vsk}) with respect to $\tau$,
%$\dot{\ds}$ can be approximately written in terms of $\ds$ as
%$|\dot{\ds}| \sim H |\ds| k \tau \ll H |\ds|$ on super horizon scales.
%Applying this relation to (\ref{ds-long}), we get $\ds \simeq$ constant on large scales after the inflationary era.

To check the validity of the above consideration, we solve (\ref{kg2}), (\ref{dot-del}), (\ref{dot-u}) and (\ref{dot-s})
numerically using the modified version of CMBEASY \cite{cmbeasy}.
In the following, the cosmological parameters for the background are chosen such that
$\Om_b^{(0)} = 0.044$, $\Om_{\rm CDM}^{(0)} = 0.216$, $\Om_d^{(0)} =0.74$, $h =0.72$.
%and the tilt of primordial curvature perturbation $n_s = 0.95$.
In our numerical integration, the form of $G_{IJ}$ is chosen according to (\ref{gij}), so that
\be
\s^1 = - \frac{\e^2 + \al \e^1}{\sqrt{1 - \al^2}} \qquad {\rm and} \qquad 
\s^2 = \frac{\e^1 + \al \e^2}{\sqrt{1 - \al^2}}\,.
\label{s12}
\ee

\begin{figure}[ht]
\includegraphics[height=0.4\textwidth, width=0.9\textwidth,angle=0]{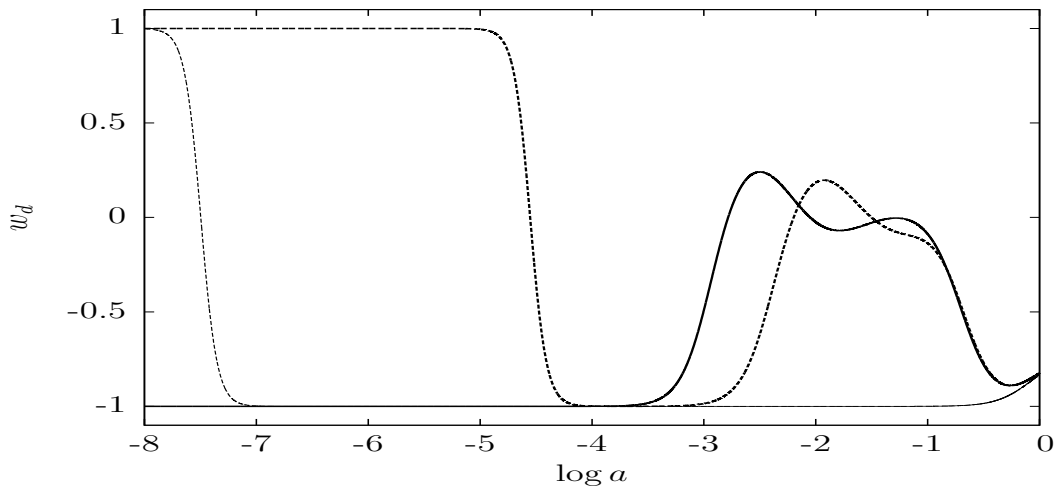}
\caption{
The evolution of $w_d$ and the equation of state parameter $w_2$ of $\phi^2$
for various initial conditions in the case of quintessence model.
The tick lines represent $w_d$, while the thin lines represent $w_2$.
In this plot, $\lm_1 = 10$, $\lm_2 = 1$ and $\al = 0$.
}
\label{fig:1}
\end{figure}

We first consider the case where $\al =0$ using the quintessence model whose Lagrangian is given by (\ref{quint}).
The evolutions of $w_d = P / \rho_d$ and $w_2$, which is the equation of state parameter of subdominant field $\phi^2$,
for various initial conditions are plotted in figure (\ref{fig:1}).
There are four types of the initial conditions in figure (\ref{fig:1}).
These four types correspond to the four possible combinations of the thick and thin lines in this figure,
e.g., the thick solid together with thin long dashed lines  correspond to the case where
dark energy is initially in the potential regime $w_d \approx -1$, while $\phi^2$ is in the kinetic regime $w_2 \approx 1$ initially.
All of these initial conditions, except the case where $w_d \approx -1$ while $w_2 \approx 1$ initially,
 lead to the same evolution of $\ds$ in the early epoch, as shown by thin solid line in figure (\ref{fig:3}),
This evolution of $\ds$ is also similar to that in the case of $\al \neq 0$ represented by thick solid line in the same figure.
In the exceptional case where $w_d \approx -1$ while $w_2 \approx 1$ initially,
the kinetic energy of $\phi^2$ can be larger than that of $\phi^1$ initially
although $\phi^2$ gives a negligible contribution to the energy density $\rho_d$, so that
$|\e^2 | > |\e^1|$ at the initial stage, and therefore the above analytical consideration is not applicable.
The evolution of $\ds$ for this case is represented by thin long dashed line in figure (\ref{fig:3}).
This figure shows that $\ds$ decreases until $w_2$ is close to $-1$, i.e., $|\e^1|$ becomes larger than $|\e^2|$.
Due to the source term on the RHS of (\ref{dot-s}), $\ds$ does not decrease to zero.
Thus, this case does not imply the vanishing of $\ds$.

\begin{figure}[ht]
\includegraphics[height=0.4\textwidth, width=0.9\textwidth,angle=0]{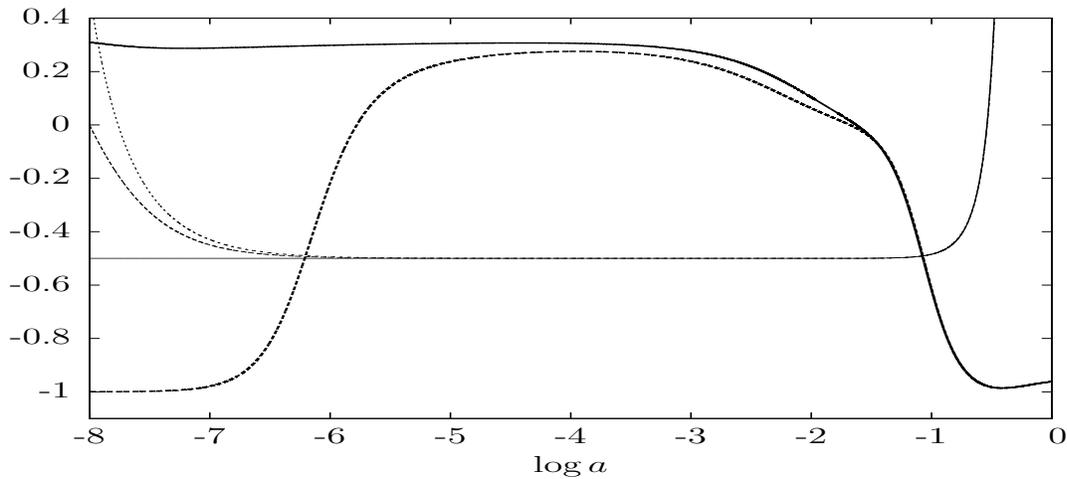}
\caption{
The evolution of $w_d$ and $\dot\phi^2/\dot\phi^1$ for the case of k-essence model.
The thick line represents the evolution of $w_d$, while the thin lines correspond to $\dot\phi^2/\dot\phi^1$.
The tin solid, thin long dashed and thin dashed lines
correspond to $\dot\phi^2= -\al\dot\phi^1$, $\dot\phi^2= 0$ and $\dot\phi^2= \al\dot\phi^1$ at initial time respectively.
In this plot, $\lm_1 = 50$, $\lm_2 = 1$ and $\al = 0.5$.
}
\label{fig:2}
\end{figure}

\begin{figure}[ht]
\includegraphics[height=0.4\textwidth, width=0.9\textwidth,angle=0]{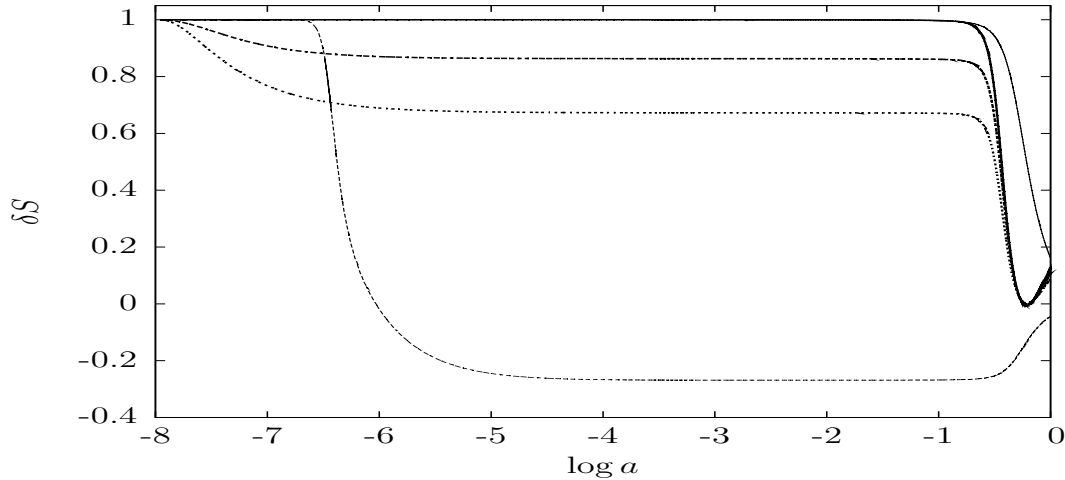}
\caption{
The evolution of the long wavelength mode of $\ds$ for models in figures (\ref{fig:1}) and (\ref{fig:2}).
The thin and thick lines correspond to the cases of $\al = 0$ and $\al = 0.5$ respectively.
The thin long dashed line is the case where $w_2 \approx 1$ while $w_d \approx -1$ initially.
The thick solid, thick long dashed and thick dashed lines
correspond to $\dot\phi^2= -\al\dot\phi^1$, $\dot\phi^2= 0$ and $\dot\phi^2= \al\dot\phi^1$ at initial time respectively.
}
\label{fig:3}
\end{figure}

We now consider the case where $\al \neq 0$ using the k-essence model whose Lagrangian is given by (\ref{kesse}).
The initial conditions for k-essence in figure (\ref{fig:2}) are chosen such that 
either $\dot\phi^2 = \pm \al \dot\phi^1$ or $\dot\phi^2 = 0$ initially,
while dark energy is either in potential regime $w_d \approx -1$ or approximately in the scaling regime $w_d\approx w_f$ at initial stage. 
In the case where dark energy starts in scaling regime,
figure (\ref{fig:2}) shows that $\dot\phi^2$ converges to $\dot\phi^2 = -\al\dot\phi^1$ rather quickly.
Although in the cases where $\dot\phi^2 = \al\dot\phi^1$ and $\dot\phi^2 = 0$ initially,
$\phi^2$ is subdominant and dark energy can be approximately in the scaling regime during the initial stage,
the relation $\dot\phi^2 \simeq -\al\dot\phi^1$ is not satified.
Hence, in these cases, $\ds$ is not constant during the initial stage as shown by thick long dashed and dashed lines in figure (\ref{fig:3}).
However, after the initial transient, $\ds$ becomes constant.
We also consider the case where $w_d \approx -1$ and $\dot\phi^2 = -\al \dot\phi^1$ initially.
In this case $\ds$ is also constant in the early epoch.
The above results for $\al \neq 0$ also hold in the case of quintessence model.

From the above consideration, it can be concluded that in the early epoch
if $\phi^2$ is subdominant compared with $\phi^1$,
the entropy perturbations in dark energy $\ds$ tend to be constant on large scales,
and consequently can survive until the present epoch for generic evolution of dark energy.
The constancy of the entropy perturbations is not due to the particular form of Lagrangian (\ref{p-assi}),
but is a consequence of $|\Xi | \ll H$ and $\mu_S^2 \ll H^2$,
which can generally occur in two-field system when one field is subdominant compared with another.

\subsection{The effects of entropy perturbations on the CMB power spectrum}

\begin{figure}[ht]
\includegraphics[height=0.4\textwidth, width=0.9\textwidth,angle=0]{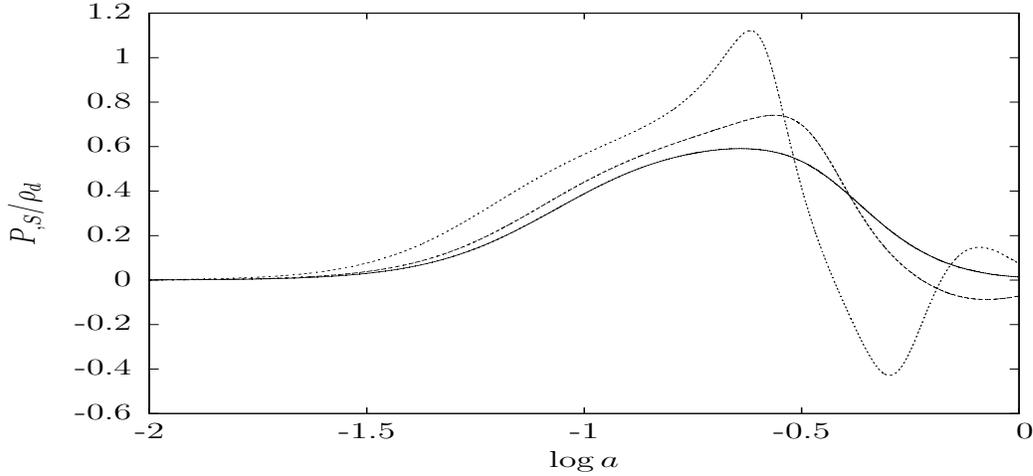}
\caption{
The evolution of $\Ps / \rho_d$. 
for various $\al$.
The solid, long dashed and dashed lines represent $\al = 0, 0.5$ and 0.8 respectively.
}
\label{fig:4}
\end{figure}

Having consider the evolution of $\ds$ in the early epoch,let us now study the evolution and effects of $\ds$ about the present era.
Since $\Xi$ and $\mu_S^2$ depend on $\Ps$ and $\PXs$,
the evolution of $\ds$ and the influence of $\ds$ on $\de_d$ are governed by these quantities.
However, since the contributions from $\Ps$ and $\PXs$ are approximately the same, sometime we consider $\Ps$ only.
As have been shown, $|\Ps / \rho_d| \ll 1$ in the early era.
Using  similar approach one expects that $|\Ps / \rho_d| < 1$ in the present epoch.
This is because when $\phi^2$ is dominant compared with $\phi^1$ at present, we have $|\e^2| > |\e^1|$, $|\s^1| > |\s^2|$
and $|P_{,2}| > |P_{,1}|$ so that 
$|\Ps | \sim | P_{,1}\s^1 + P_{,2}\s^2| < |P_{,2}\e^2| \sim |P_{,\sig}| \sim \rho_d$.
The ratio $|\Ps / \rho_d|$ can be of order of unity during the transition from $\phi^1$ - dominant to $\phi^2$ - dominant,
because $|\s^2|$ and $|P_{,2}|$ are of order of $|\s^1|$ and $|P_{,1}|$ respectively.
This rough estimate is confirmed by the numerical results in figure (\ref{fig:4}).
Hence, $\ds$ can have a contribution to the dynamics of $\de_d$ only during the transition stage,
because the contribution from the entropy perturbations to the pressure perturbations in (\ref{dp-ds}) depends on the magnitude of $\Ps$ and $\PXs$.
This implies that in the general two-field dark energy models,
although the entropy perturbations in dark energy can be constant until the present epoch
because one field is subdominant compared with the other in the early era,
the entropy perturbations cannot significantly influence the dynamics of density perturbations in the universe
unless the subdominant field becomes more important at late time.

\begin{figure}[ht]
\includegraphics[height=0.4\textwidth, width=0.9\textwidth,angle=0]{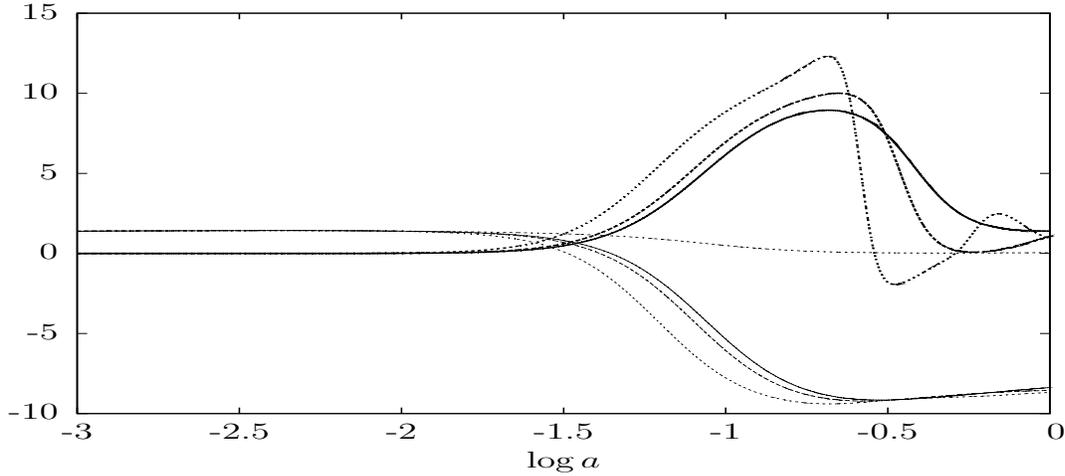}
\caption{
The evolution of $\tilde{\Ps}\ds$ (thick lines) and $\de_d$ (thin lines).
the solid, long dashed and dashed lines correspond to $\al =0$, $0.5$ and $0.8$ respectively.
For comparison , the evolution of $\de_d$ in the case where $\ds = 0$ is plotted using dotted line.
In this plot, $k \simeq 2\times 10^{-4}$ Mpc$^{-1}$.
}
\label{fig:5}
\end{figure}

The evolution of $\ds$ can influence the evolution of $\de_d$ via the relation $\de P/\rho_d \propto \tilde{\Ps}\ds$ given in (\ref{dp-ds}).
Here, $\tilde{\Ps} = \[\Ps\(1 + \ce^2\) - 2\ce^2 X \PXs\]/\rho_d$.
We follow CMBEASY code to normalize all perturbation variables such that ${\cal R} = -1$.
If the value of $\tilde{\Ps}\ds$ is sufficiently large and possitive, $\ds$ will give a possitive contribution to $\de P$,
and also $\dot u_d$ in (\ref{dot-u}).
Consequently, it follows from (\ref{dot-del}) that $\de_d$ can decrease more rapidly on large scales when $\tilde{\Ps}\ds$ increases.
This qualitative consideration is independent of the form of $P(X,\phi^I)$ and
in agreement with  the numerical results in figure (\ref{fig:5}).
In contrast, if $\tilde{\Ps}\ds$ is sufficiently large and negative, its contributions should increase $\dot{\de_d}$ on large scales.

For illustration, in figures (\ref{fig:4}) and (\ref{fig:5}), we plot the evolutions of $\Ps/\rho_d$, $\tilde{\Ps}\ds$ and $\de_d$
from the numerical integration for k-essence model whose Lagrangian is given by (\ref{kesse}).
Since the evolutions of these quantities for our choice of quintessence and k-essence models have approximately the same features,
we consider the case of k-essence model only.

We set $\lm_1 = 50$ and $\lm_2 =1$ in the numerical integration.
The initial conditions for the background field are chosen such that dark energy is in the scaling regime
while $\dot\phi^2 = -\al\dot\phi^1$ initially.
For the perturbed quantities, the adiabatic initial conditions given in (\ref{adia-con}) are used for $\de_d$ and $u_d$.
The initial value of $\ds$ are chosen such that the CMB power spectrum at $\ell =2$ for $\al =0$ is maximally suppressed.
This initial value can be in agreement with the initial value from inflation if $|\PX | < 1$ during inflation.
In our case, initial $\ds / {\cal R} < 0$ corresponds to
$\tilde{\Ps}\ds / {\cal R} < 0$ at the initial time, and leads to $\tilde{\Ps}\ds / {\cal R} < 0$ during the transition stage.
In general, the relation between the sign of $\tilde{\Ps}\ds$ during the transition stage and
the relative signs of $\ds$ or $\tilde{\Ps}\ds$ to ${\cal R}$ at the initial time depends on the form of  Lagrangian for dark energy.

Since the effects of entropy perturbations in dark energy are most important on large scales 
we consider the perturbation mode whose wavenumber is $k \simeq 2\times 10^{-4}$ Mpc${}^{-1}$.
This mode enters the Hubble radius about the present epoch.
%
%
%Figure (\ref{fig:4}) shows that when the transition stage begins $\Ps / \rho_d$ starts to increase.
%The rate of increase and the maximum amplitude of $\Ps / \rho_d$ increase with $\al$.
%Nevertheless, when $\al$ increases, $\Ps /\rho_d$ decreases more rapidly after it reaches its maximum.
%
%It can be seen from figure (\ref{fig:5})
%that the evolution of $\tilde{\Ps}\ds$ has approximately a similar feature as $\Ps / \rho_d$.
%This is because $\tilde{\Ps} \sim \Ps / \rho_d$, and
%after $\Ps / \rho_d$ reaches it first maximum, the evolution of $\ds$ is governed by $\tilde{\Ps}$, such that $\ds$ evolves approximately in a similar 
%manner to $\tilde{\Ps}$.
%

The effects of entropy perturbations on the density perturbations in the universe can be studied using the relation between
the metric perturbations and $\de_d$ in (\ref{ein1}).
Since we are interested in the perturbations on large scales
and radiation contributes a negligible fraction of the total energy density during the transition stage,
we write (\ref{ein1}) as
\be
\dot\Phi = - H\Phi
- \frac H2 \(\Om_m\de_m + \Om_d\de_d\)\,,
\label{dot-phi}
\ee
where the subscript $m$ denotes matter.
The relation $\Phi \sim \Psi$ has been used in the above equation because radiation is negligible.
Hence, if the contributions from density perturbations is ignored for a while,
it can be read from (\ref{dot-phi}) that the amplitude of metric perturbations decreases with time due to the expansion of the universe.
The decreases of $|\Phi|$ and $|\Psi|$ which correspond to the ISW effect,
partially cancel the ordinary Sachs-Wolfe (OSW) effect.
It follows from (\ref{dot-phi}) that the inclusion of density perturbations can modify the evolution of $\Phi$,
such that if $\Phi / \de_d > 0$, $|\Phi|$ will decrease faster, and $|\Phi|$ will decrease more slowly or increase if $\Phi /\de_d < 0$
\cite{GorHu:04, Moroi:04, kk:07}.
When $|\Phi|$ decreases faster, the ISW contributions will be larger, leading to more cancellations between ISW and OSW effects.
In this case, the CMB power spectrum is suppressed  at low multipoles.
In contrast, OSW effect is less canceled by ISW effect when $|\Phi|$ decreases more slowly,
and OSW and ISW can add up when $|\Phi|$ increases.
Consequently, the CMB power spectrum is enhanced  at low multipoles in these cases.

\begin{figure}[ht]
\includegraphics[height=0.7\textwidth, width=0.9\textwidth,angle=0]{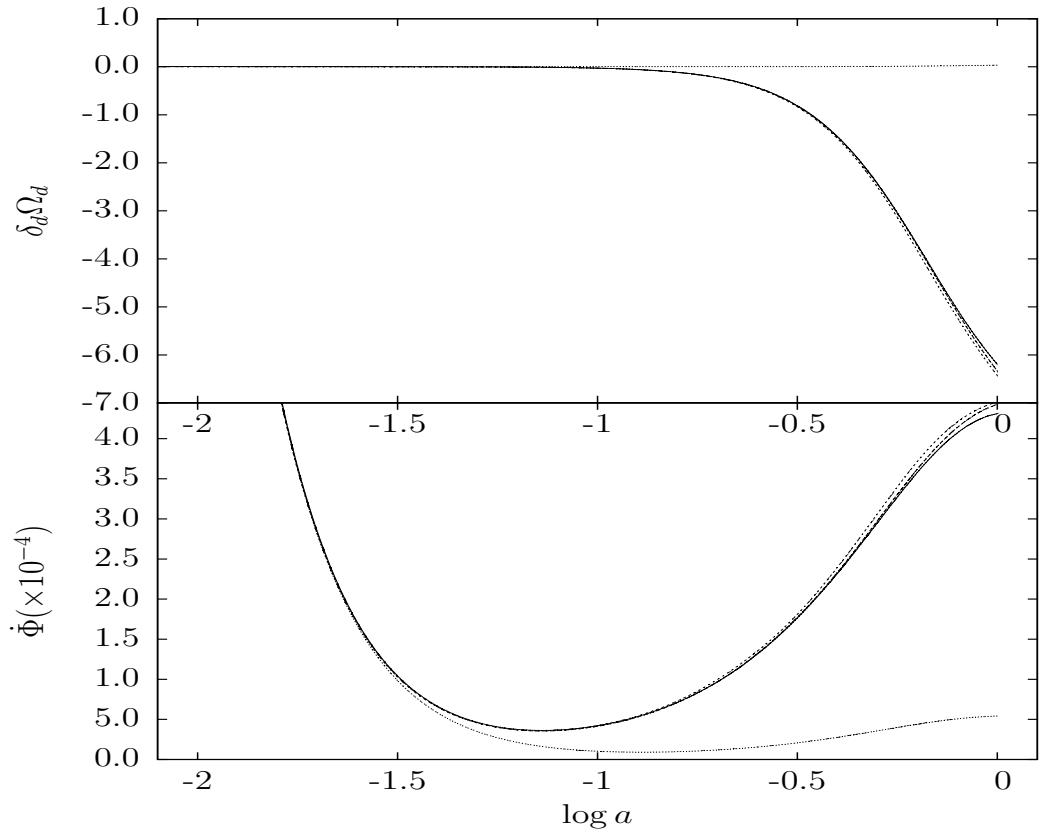}
\caption{
The evolution of $\Om_d\de_d$ is shown in the upper panel, while
the evolution of $\dot\Phi$  is shown in the lower panel.
In the both panels, the solid, long dashed and dashed lines correspond to $\al =0$, $0.5$ and $0.8$ respectively.
The evolution of both quantities  in the case where $\ds = 0$ is denoted by dotted line.
In this plot, $k \simeq 2\times 10^{-4}$ Mpc$^{-1}$.
}
\label{fig:6}
\end{figure}

\begin{figure}[ht]
\includegraphics[height=0.4\textwidth, width=0.9\textwidth,angle=0]{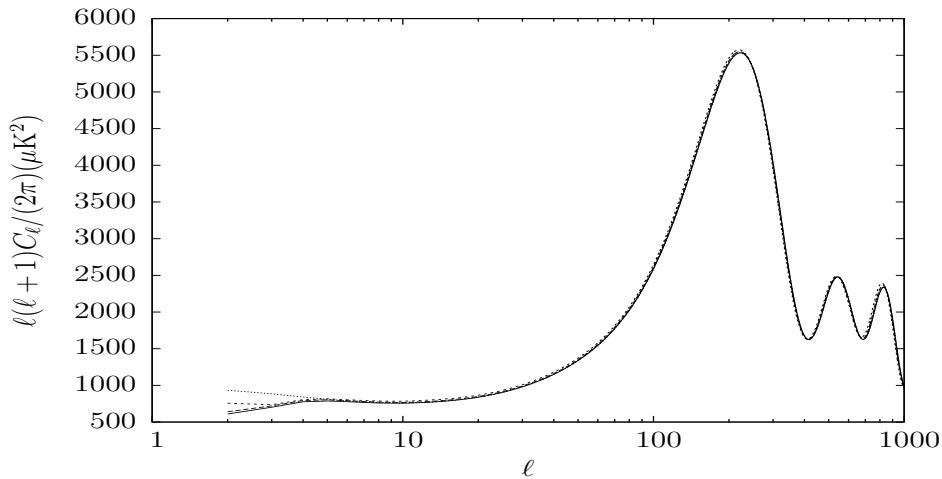}
\caption{
The CMB power spectra for the k-essence model.
The solid, long dashed and dashed lines correspond to $\al = 0$, $0.5$ and $0.8$ respectively.
The case where $\ds =0$ while $\al =0$ is denoted by dotted line.
}
\label{fig:7}
\end{figure}

Using the same model of k-essence and initial conditions as above,
we plot the evolutions of $\Om_d\de_d$ and $\dot\Phi$ in figure (\ref{fig:6})
and plot the CMB power spectrum in figure (\ref{fig:7}).
In our numerical integration, $\Phi$ is negative because all perturbation variables are normalized such that ${\cal R} = -1$.
Thus, in figure (\ref{fig:6}), $\dot\Phi$ increases when $\Om_d\de_d < 0$.
This means that due to the contribution from $\de_d$, $|\phi|$ decreases more rapidly, i.e., the ISW effect is enhanced.
The enhancement of ISW effect can lead to the suppression of the CMB power spectrum
at low multipoles as shown in figure (\ref{fig:7}).
These are in agreement with the above qualitative consideration

From the above analysis and figures (\ref{fig:5}) - (\ref{fig:7}),
one can conclude that if the curvature perturbation ${\cal R}$ and the contribution from entropy perturbations
in dark energy, which is proportional to $\tilde{\Ps}\ds$, have opposite signs after the matter era,
the ISW effect can be enhanced and consequently the CMB power spectrum can be suppressed at low multipoles.
In contrast, if ${\cal R}$ and $\tilde{\Ps}\ds$ have the same signs after the matter era, the CMB power spectrum at low multipoles can be enhanced.
This is in agreement with literature \cite{GorHu:04, Moroi:04}.
However, for two-field dark energy models, the contribution from entropy perturbations in dark energy depends on $\tilde{\Ps}$,
i.e., the evolution of dark energy, during the transition from $\phi^1$ - dominant to $\phi^2$ - dominant.
Hence, in general, $\tilde{\Ps}\ds / {\cal R} < 0$ at the initial time does not necessarily lead to
$\tilde{\Ps}\ds / {\cal R} < 0$ after the matter era, and consequently the suppression of the CMB power spectrum at low multipoles.

\begin{figure}[ht]
\includegraphics[height=0.4\textwidth, width=0.9\textwidth,angle=0]{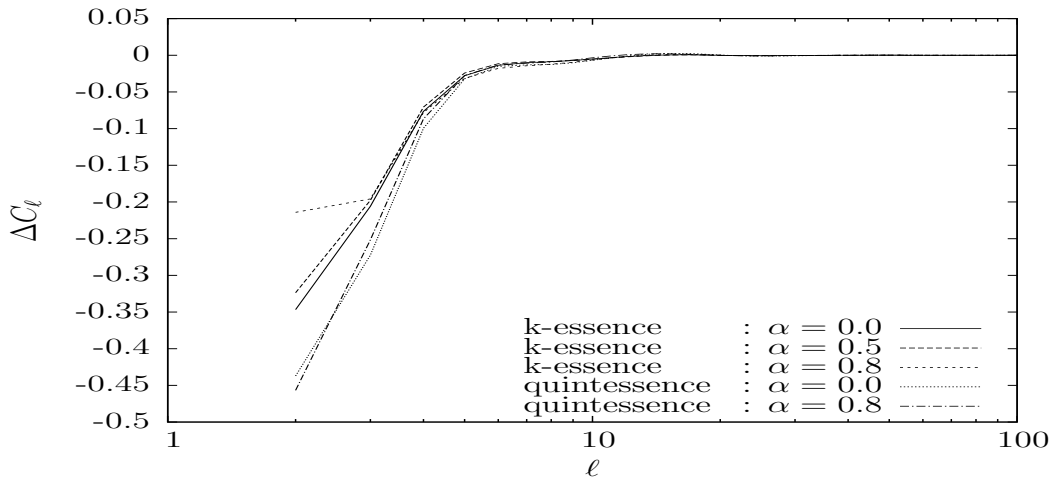}
\caption{
The fractional difference in the CMB power spectra for k-essence and quintessence models.
}
\label{fig:8}
\end{figure}

In order to study the influence of $\al$ on the effect of entropy perturbations, we plot
the fractional difference in the CMB power spectra for k-essence and quintessence in figure (\ref{fig:8}).
The fractional difference in the CMB power spectra between  
the case of $\ds = 0$ and the case of $\ds \neq 0$ is expressed as the ratio
\be
\De C_\ell = \frac{C_\ell - C_\ell^*}{C_\ell^*}\,,
\ee
where $C_\ell^*$ is the power spectrum for the case of $\ds = 0$ and
$C_\ell$ is the power spectrum for the case of $\ds \neq 0$.

In this figure, the initial value for $\ds$ in the case of quintessence
is also chosen such that the CMB power spectrum at $\ell =2$ for $\al =0$ is maximally suppressed.
This initial value of $\ds$ is much larger than a possible initial value from inflation  given in (\ref{init-ds}),
because $\PX =1$ for quintessence. If the initial value in (\ref{init-ds}) is used,
the entropy perturbations have small contribution to ISW effect and hence have no effect on CMB power spectrum.
This figure shows that the suppression of the CMB power spectrum depends on $\al$ in the case of k-essence,
but weakly depends on $\al$ in the case of quintessence.
The influence of $\al$ on the suppression of CMB power spectrum strongly depends on the form of $g(Y)$ in (\ref{p-assi}).
%This implies that the effects of entropy perturbations depend on the evolution of $\tilde{\Ps}\ds$ during the transition stage.
%

\section{Conclusion}

In this work, we study dynamics of entropy perturbations in two-field assisted dark energy model.
Based on the scenario for this dark energy model in which one field is subdominant compared with the other in the
early era, we show that the entropy perturbations in this two-field system tend to be constant on large scales in the early era,
and hence can survive until the present epoch, for a generic evolution of dark energy during the radiation and matter eras.
This behaviour of the entropy perturbations is preserved even when the fields are coupled via kinetic interaction.
Although in our case the constancy of the entropy perturbations is obtained using the Lagrangian which
has scaling behaviour, this result can also be obtained from more general form of Lagrangian.
This is because the constancy of the entropy perturbations is the consequence of the subdomination
of one field compared with the other in the early epoch.
However,  the entropy perturbations cannot significantly influence the dynamics of density perturbations in the universe
unless the subdominant field in the early epoch becomes more important at late time.

Since, for assisted dark energy, the subdominant field in the early epoch becomes dominant at late time,
the entropy perturbations can influence the dynamics of metric perturbations so that the ISW effect is modified.
The influences of the entropy perturbations on the ISW effect depend  on both the initial condition for $\ds$ and the evolution of dark energy during 
the transition between $\phi^1$- domination and $\phi^2$- domination after the matter era.
Assuming correlations between the entropy and curvature perturbations,
the ISW effect will be enhanced and therefore the CMB power spectrum will be suppressed at low multipoles
if the contribution from entropy perturbations and the curvature perturbations have opposite signs,
i.e., $\tilde{\Ps}\ds/{\cal R} < 0$, after the matter era.
For our choice of models in the numerical integration,
the negative  $\tilde{\Ps}\ds/{\cal R}$ at initial time leads to negative $\tilde{\Ps}\ds/{\cal R}$ after the matter era.
However, in general, initial $\tilde{\Ps}\ds/{\cal R} < 0$ does not necessarily imply $\tilde{\Ps}\ds/{\cal R} < 0$ after the matter era.

Assuming dark energy  to exist since inflationary epoch,
the initial value of $|\ds |$ estimated during inflationary epoch cannot be large enough to influence the ISW effects in the case of quintessence.
This is because $\PX =1$ for usual quintessence models.
The initial value of $|\ds |$ can be large enough to influence the ISW effect if $|\PX |$ is significantly smaller than unity
during inflationary era, which is possible in many k-essence models.

If dark energy fields are coupled via kinetic terms, the modifications of CMB power spectrum due to entropy perturbations
also depend on the coupling strength $\al$.
For our choice of models in the numerical integration,
this dependence is significant in the case of k-essence, but not in the case of quintessence.
The influence of $\al$ on the CMB power spectrum strongly depends on the form of $g(Y)$ in (\ref{p-assi}).

\section*{Acknowledgments}

The author would like to thank A. Ungkitchanukit and anonymous referee for comments on the manuscript.
This work is supported by Thailand Research Fund (TRF) through grant TRG5280036.

\end{document}